\documentclass[a4paper]{article}

\usepackage{INTERSPEECH_v2}
\usepackage{amssymb,amsmath,bm,url}
\usepackage{textcomp}
\usepackage{multirow}
\usepackage{ctable}

\title{On the use of DNN Autoencoder for Robust Speaker Recognition}
\name{Ond\v{r}ej Novotn\'y, Old\v{r}ich Plchot, Pavel Mat\v{e}jka, Ond\v{r}ej Glembek}
\address{Brno University of Technology, Speech@FIT and IT4I Center of Excellence, Czechia
\thanks{The work was supported by Czech Ministry of Interior project No. VI20152020025 "DRAPAK", Google Faculty Research Award program, Czech Science Foundation under project No. GJ17-23870Y, and by Czech Ministry of Education, Youth and Sports from the National Programme of Sustainability (NPU II) project "IT4Innovations excellence in science - LQ1602".}
}
\email{\{inovoton,iplchot,matejkap,glembek\}@fit.vutbr.cz}
\begin{document}

\maketitle
\begin{abstract}
In this paper, we present an analysis of a DNN-based autoencoder for speech enhancement, dereverberation and denoising. The target application is a robust speaker recognition system. We started with augmenting the Fisher database with artificially noised and reverberated data and we trained the autoencoder to map noisy and reverberated speech to its clean version. 
We use the autoencoder as a preprocessing step for a state-of-the-art text-independent speaker recognition system. We compare results achieved with pure autoencoder enhancement, multi-condition PLDA training and their simultaneous use. We present a detailed analysis with various conditions of NIST SRE 2010, PRISM and artificially corrupted NIST SRE 2010 telephone condition. We conclude that the proposed preprocessing significantly outperforms the baseline and that this technique can be used to build a robust speaker recognition system for reverberated and noisy data.

\end{abstract}

\section{Introduction}
In last years, various techniques for speech and signal processing have been introduced to cope with the distortions caused by noise and reverberation. 
In the field of speaker recognition, one way to tackle this problem is to use multi-condition training of PLDA, where we introduce noise variability and reverberation
variability into the within-class variability of speakers. 
Also, several techniques were introduced in the field of microphone array to solve this issue by active noise canceling, beamforming and filtering~\cite{kumatani:micarray:2012}.
For single microphone systems, front-ends utilize signal pre-processing methods such as Wiener filtering, adaptive voice activity detection (VAD), gain control, etc.~\cite{ETSI:07}.
Next, various designs of robust features~\cite{plchot:rats13} are used in combination with normalization techniques such as cepstral mean and variance normalization or short-time gaussianization~\cite{Pelecanos01}.

The last years have seen, the rise of interest in NN signal pre-processing. An example of classical approach to remove a room impulse response is proposed in \cite{Dufera2009}, where the filter is estimated by an NN. NNs have also been used for speech separation in \cite{Yanhui2014}. NN-based autoencoder for speech enhancement was proposed in \cite{Xu2014} with optimization in
\cite{Xu2014a} and finally, reverberant speech recognition with signal
enhancement by a deep autoencoder was tested in the Chime Challenge
and presented in \cite{Mimura2014}.  

In this paper, we investigate the use of a DNN autoencoder as an audio pre-processing front-end for speaker recognition. The autoencoder is trained to learn a mapping from noisy and reverberated speech to clean speech. The frame-by-frame aligned examples for DNN training are artificially created by adding noise and reverberation to the Fisher speech corpus.  The analysis in this paper extends our previous work presented in \cite{plchot-enhancement-2016} and focuses on different autoencoders in more variable and harder conditions. 
These conditions are simulated by adding the noise and reverberation into the NITST SRE2010 telephone condition and extend the selection of test sets that we used in \cite{plchot-enhancement-2016}. 

We confirm our conclusions from \cite{plchot-enhancement-2016} and we offer more experimental evidence and thorough analysis to demonstrate that the proposed method increases the performance of the text independent speaker recognition system. As it was already shown that performing multi-condition training with added noisy and reverberated data helps significantly in speaker recognition~\cite{david:icassp:vts,lei:multistyle}, we will also discuss the influence of quantity, quality, and type of autoencoder training data on performance of the analyzed SRE system.
In the end, we will show that we can significantly profit from combination of both techniques.

\section{Autoencoder training and dataset design}
\label{data}
Fisher English database parts 1 and 2 were used for training the autoencoder. 
They contain over 20,000 telephone conversational sides or approximately 
1800 hours of audio.

Our autoencoder consists of three hidden layers with 1500 neurons in each layer.  The input of the autoencoder was central frame of a log-magnitude spectrum with context of +/- 15 frames (in total 3999-dimensional input). The output is an 129-dimensional enhanced central frame. We used Mean Square Error (MSE) as objective function during training.

\subsection{Adding noise}
We prepared a noise dataset that consists of three sources of different types of noise:
\begin{itemize}
\item 272 samples (4 minutes long) taken from the Freesound library \footnote{\url{http://www.freesound.org}} (real fan, HVAC, street, city, shop, crowd, library, office and workshop).
\item 7 samples (4 minutes long) of artificially generated noises: various spectral modifications of white noise + 50 and 100 Hz hum.
\item 25 samples (4 minutes long) of babbling noises by merging speech from 100 random speakers from Fisher database using speech activity detector. 
\end{itemize}

\noindent
Noises were divided into three disjoint groups for training (223 files), development (40 files) and test (41 files). 

\subsection{Reverberation}
We prepared two sets with room impulse responses (RIRs). 
The first set consists of real room impulse responses from several databases: AIR \cite{AIR:WWW}, C4DM \cite{C4DM:WWW,C4DM:WWW2}, MARDY \cite{MARDY:WWW}, OPENAIR \cite{OpenAir:WWW}, RVB 2014 \cite{RVB:WWW}, RWCP \cite{RWCP:WWW}. 
Together, they form a set with all types of rooms (small rooms, big rooms, lecture room, restrooms, halls, stairs etc.). All room models have more than one impulse response per room (different RIR was used for source of the signal and source of the noise to simulate different locations of their sources). Rooms were split into two disjoint sets, with 396 rooms for training, 40 rooms for test.

The second set consists of artificially generated room impulse responses using ``Room Impulse Response Generator" tool from E. Habets~\cite{Habet:rir}.
The tool can model the size of room (3 dimensions), reflectivity of each wall, type of microphone, position of source and microphone, orientation of microphone towards the audio source,  and number of bounces (reflections) of the signal.
We generated a pair of RIRs for each room model (one used for source of the sound, one for source of the noise).  Again we generated two disjoint sets, with 1594 RIRs for training and 250 RIRs for test.

\subsection{Composition of the training set}
\label{section:signal_corruption}
\begin{figure}[tb]
    \begin{center}
      \scalebox{0.5}{\includegraphics{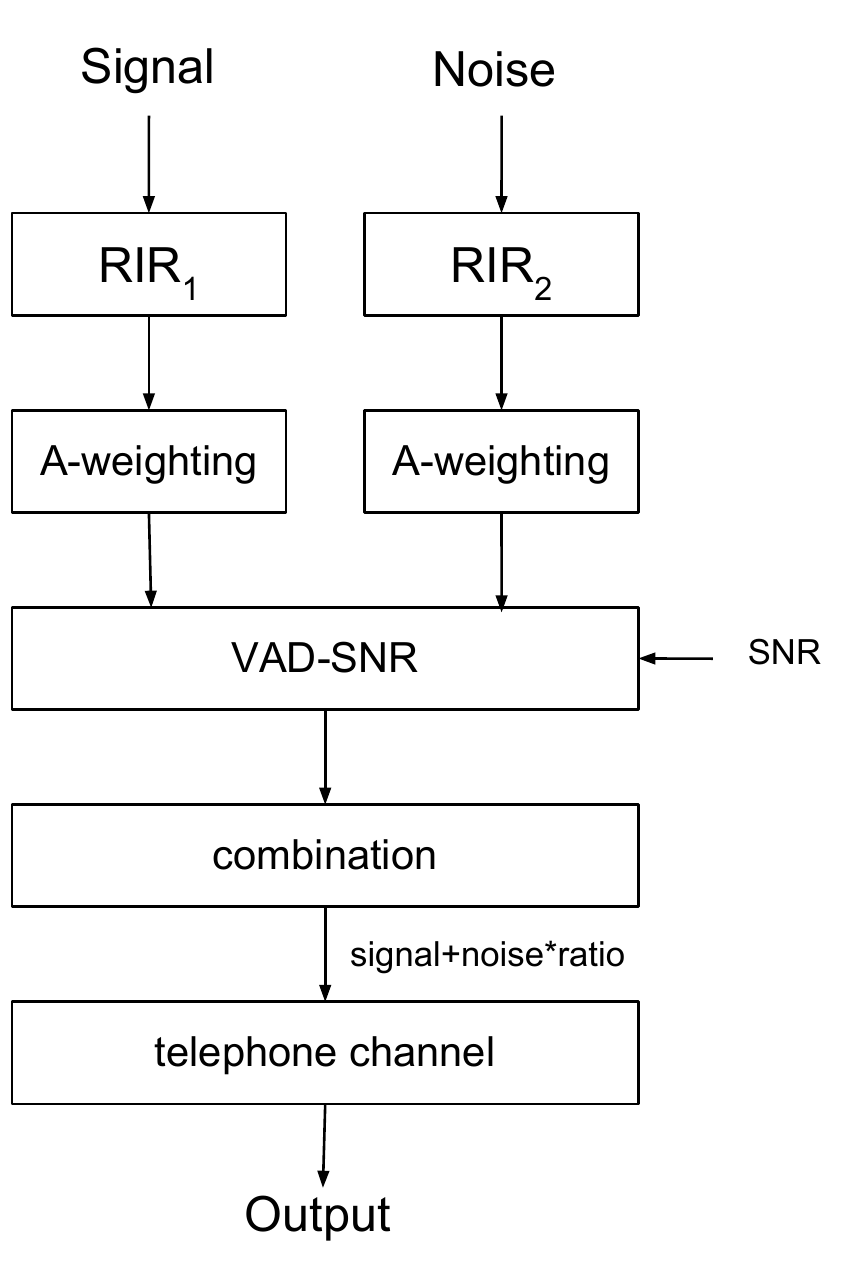}}
    \end{center}
    \vspace{-5mm}
    \caption{The process of data preparation (corruption) for autoencoder training or new SRE condition design.}
    \label{fig:noising_pipeline}
\end{figure}

To mix the reverberation, noise and signal at given SNR, we followed the  procedure showed in figure \ref{fig:noising_pipeline}.
The pipeline begins with two branches, when speech and noise are reverberated separately. Different RIRs from the same room are used for signal and noise, to simulate different positions of sources.

The next step is A-weighting.  A-weighting is applied to simulate the perception of the human ear to added noise \cite{aweighting}. With this filtering, the listener would be able to better perceive the SNR, because most of the noise energy is coming from frequencies, that the human ear is sensitive to. 

In the following step, we set a ratio of noise and signal energies to obtain the required SNR. Energies of the signal and noise are computed from frames given by original signal's voice activity detection (VAD). It means the computed SNR is really present in speech frames which are important for our recognition (frames without voice activity are removed during processing).

After the combination, where signal and noise are summed together at desired SNR, we filter the resulting signal with telephone channel. To compensate for the fact that our noise samples are not coming from the telephone channel, while the original clean data (Fisher, NIST tel-tel) are in fact telephone.
The final output is a reverberated and noisy signal with required SNR, which simulates a recording passing through the telephone channel (as was the original signal) in various acoustic environments. 
In case we want to add only noise or reverberation, the appropriate part of the algorithm is used. 

\section{Speaker recognition system}
Our systems are based on i-vectors~\cite{DehakN_TASLP:2010,PLDA:kenny}.
To train i-vector extractors, we always use 2048-component diagonal-covariance Universal Background Model (GMM-UBM) and we set the dimensionality of i-vectors to 600. 
We apply LDA to reduce the
dimensionality to 200.  Such processed i-vectors are then transformed by global
mean normalization and length-normalization~\cite{DehakN_TASLP:2010,RomeroD_ICSLP:2011}.  

Speaker verification score is produced by comparing two i-vectors
corresponding to the segments in the verification trial by means of PLDA~\cite{PLDA:kenny}.

In our experiments, we used cepstral features, extracted using a 25\,ms Hamming
window.  We used 24 Mel-filter banks and we limited the bandwidth to the
120--3800Hz range.  19~MFCCs together with
zero-\emph{th} coefficient were calculated every 10\,ms. This
20-dimensional feature vector was subjected to short time mean- and
variance-normalization using a 3\,s sliding window.  Delta and double delta
coefficients were then calculated using a five-frame window giving a
60-dimensional feature vector. 

After feature extraction, voice activity detection (VAD) was performed by the 
BUT Czech phoneme recognizer~\cite{matejka-odyssey-2006}, dropping all frames 
that are labeled as silence or noise.  The recognizer was trained on the Czech 
CTS data, but we have added noise with varying SNR to 30\% of the database.

\subsection{Datasets}
We used the PRISM~\cite{ferrer:sre11} training dataset definition without added 
noise or reverb to train UBM and i-vector transformation. 
Five variants of gender independent PLDA were trained: one only on the clean training data, the rest 
included also artificially added different cocktail of noises and reverb. 
Artificially added noise and reverb segments totaled approximately twenty-four thousand segments or $30\%$ of total number of clean segments for PLDA training.  
The PRISM set comprises Fisher 1 and 2, Switchboard phase 2 and 3 and Switchboard cellphone phases 
1 and 2, along with a set of Mixer speakers. This includes the 66 held out 
speakers from SRE10 (see Section III-B5 of~\cite{ferrer:sre11}), and 965, 980, 485 and 310 speakers 
from SRE08, SRE06, SRE05 and SRE04, respectively. A total of 13,916 speakers 
are available in Fisher data and 1,991 in Switchboard data.

We evaluated our systems on the \emph{female} portions of the following conditions 
in NIST SRE 2010~\cite{NIST_SRE:WWW} and PRISM~\cite{ferrer:sre11}:
\begin{itemize}
 \item {\bf tel-tel}: SRE 2010 extended telephone condition 
involving normal vocal effort conversational telephone speech in enrollment and 
test (known as condition 5).
  \item {\bf int-int}: SRE 2010 extended interview condition involving 
interview speech from different microphones in enrollment and test (known as 
condition 2).
  \item {\bf int-mic}: SRE 2010 extended interview-microphone condition 
involving interview enrollment speech and normal vocal effort conversational 
telephone test speech recorded over a room microphone channel (known as 
condition 4).
 \item {\bf prism,noi}: Clean and artificially noised waveforms from both interview and telephone conversations recorded over lavalier microphones.
 Noise was added  at different SNR levels and recordings tested against each other.
 \item {\bf prism,rev}: Clean and artificially reverberated waveforms from both interview and telephone conversations recorded over lavalier microphones.
Reverberation was added with different RTs and recordings tested against each other.
\item {\bf prism,chn}: English telephone speech with normal vocal effort 
recorded over different microphones from both SRE2008 and 2010 tested against each other.
\end{itemize}

Additionally, we created new artificially corrupted evaluation sets from the NIST 2010 tel-tel condition. The process was the same as described in section \ref{section:signal_corruption} while using the tests portion of our noise and reverberation sets.  We created seven new conditions:
\begin{itemize}
\item {\bf rev-tel-tel}: SRE 2010 tel-tel condition corrupted by real room impulse responses (reverberation). 
\item {\bf noi-$\ast$-tel-tel}: SRE 2010 tel-tel condition corrupted by noise. We used three ranges of noise: 0-7dB, 7-14dB, 14-21dB (range is writen on position of $\ast$, e.g. noi-0-7-tel-tel).
\item {\bf rev-noi-$\ast$-tel-tel}:  SRE 2010 tel-tel condition corrupted by noise and real rooms impulse responses. Again, we used three ranges of noise: 0-7dB, 7-14dB, 14-21dB.
\end{itemize}
  
The difference between these new conditions and the conditions based on the PRISM set is in more realistic reverberation.  Condition \textbf{prism,rev} is created from clean microphone data corrupted with artificially generated RIRs. The new conditions focus on adding a real reverberation to the telephone data. Similarly, the \textbf{prism,noi} condition is created from microphone data by adding the noise at three levels of SNR (8dB, 15dB, 20dB), the new conditions use telephone data and randomly chosen SNR levels from the given intervals. Additionally, the selected telephone data tend to be more difficult than the microphone data used in the PRISM conditions.   

The recognition performance is evaluated in terms of the equal error rate 
(EER).

\begin{table*}[!htb]
\centering
\caption{\label{tab:textindep} Results (EER $[\%]$) obtained in four scenarios. The first two blocks correspond to the system trained only with clean data (PLDA trained on clean data). In the left block, scores of baseline system are displayed. In the right block, the score of the clean system with enhancement data is displayed. 
Results of five autoencoders trained on: N - noise, (A/R)R- artificial/real reverberation, or both ($+$) are presented in each column. 
The last two blocks correspond to systems trained in multi-condition fashion (with noised and reverberated data in PLDA). 
Results in each column correspond to different PLDA multi-condition training set: 
N - noise, (A/R)R- artificial/real reverberation, or both ($+$). The very last block present results of the combination of both techniques. For combination, we select autoencoder trained on noised and reverberated data with real reverberation (N$+$RR).}
\scalebox{0.7}{
\begin{tabular}{ l  || c ||c c c c c| c c c c c | c c c c c }\toprule
\addlinespace[0.05cm]
\multicolumn{2}{c}{} & \multicolumn{5}{c}{PLDA trained on \textbf{clean} data} & \multicolumn{10}{c}{PLDA trained on \textbf{multi-condition} data} \\ 

\addlinespace[0.05cm] 
\cmidrule(rl){2-7} \cmidrule(rl){8-17} 
\multicolumn{1}{c}{} & \multicolumn{1}{c}{baseline} & \multicolumn{5}{c}{Autoencoder training} & \multicolumn{5}{c}{PLDA extension data} & \multicolumn{5}{c}{Autoencoder (N$+$RR) + PLDA extension data} \\ 
\cmidrule(rl){2-2} \cmidrule(rl){3-7} \cmidrule(rl){8-12} \cmidrule(rl){13-17}

Condition & & N        & AR       & N$+$AR   & RR      & N$+$RR  & N & AR  & N$+$AR& RR & N$+$RR & N & AR & N$+$AR & RR & N$+$RR \\ 
\cmidrule(rl){1-17}
 tel-tel               & $2.062$ &  $2.075$ &  $2.093$ &  $2.074$ &  $\boldsymbol{1.999}$ &  $2.063$ &  $2.458$ &  $2.071$ &  $2.728$ &  $\boldsymbol{2.035}$ &  $2.796$ &  $2.480$ &  $2.070$ &  $2.677$ &  $2.143$ &  $2.752$  \\ 
 prism,noi             & $2.950$ &  $\boldsymbol{2.122}$ &  $2.497$ &  $2.256$ &  $2.470$ &  $2.190$ &  $\boldsymbol{2.265}$ &  $3.080$ &  $2.518$ &  $2.926$ &  $2.456$ &  $\boldsymbol{1.969}$ &  $2.243$ &  $2.037$ &  $2.236$ &  $2.059$  \\ 
 prism,rev             & $2.071$ &  $1.748$ &  $1.621$ &  $1.608$ &  $\boldsymbol{1.511}$ &  $1.559$ &  $2.220$ &  $\boldsymbol{1.537}$ &  $1.620$ &  $1.613$ &  $1.632$ &  $1.583$ &  $\boldsymbol{1.419}$ &  $1.385$ &  $1.422$ &  $\boldsymbol{1.419}$  \\
 int-int               & $1.756$ &  $1.792$ &  $1.693$ &  $1.766$ &  $\boldsymbol{1.634}$ &  $1.790$ &  $1.860$ &  $\boldsymbol{1.677}$ &  $1.760$ &  $1.669$ &  $1.714$ &  $1.806$ &  $\boldsymbol{1.697}$ &  $1.714$ &  $1.705$ &  $1.761$  \\
 int-mic               & $1.089$ &  $1.136$ &  $1.085$ &  $1.151$ &  $\boldsymbol{1.010}$ &  $1.112$ &  $1.226$ &  $\boldsymbol{0.770}$ &  $0.921$ &  $0.960$ &  $1.042$ &  $0.981$ &  $1.000$ &  $\boldsymbol{0.848}$ &  $0.982$ &  $0.943$  \\
 prism,chn             & $0.795$ &  $0.523$ &  $0.599$ &  $\boldsymbol{0.402}$ &  $0.596$ &  $0.428$ &  $1.000$ &  $\boldsymbol{0.544}$ &  $0.666$ &  $0.630$ &  $0.756$ &  $0.456$ &  $0.344$ &  $0.371$ &  $\boldsymbol{0.277}$ &  $0.400$  \\
 rev-tel-tel           & $19.373$ &  $14.760$ &  $11.182$ &  $13.450$ &  $\boldsymbol{9.149}$ &  $9.365$ &  $17.835$ &  $9.461$ &  $10.151$ &  $\boldsymbol{5.246}$ &  $6.598$ &  $8.287$ &  $6.137$ &  $5.847$ &  $\boldsymbol{4.066}$ &  $4.761$  \\  
 noi-14-21-tel-tel     & $4.959$ &  $\boldsymbol{3.298}$ &  $4.009$ &  $3.943$ &  $3.721$ &  $3.703$ &  $\boldsymbol{2.901}$ &  $4.605$ &  $3.530$ &  $4.321$ &  $3.390$ &  $\boldsymbol{2.689}$ &  $3.205$ &  $2.962$ &  $3.021$ &  $2.980$  \\
 noi-7-14-tel-tel      & $8.291$ &  $\boldsymbol{5.117}$ &  $6.808$ &  $5.710$ &  $6.660$ &  $5.749$ &  $\boldsymbol{3.941}$ &  $8.026$ &  $4.920$ &  $7.540$ &  $4.715$ &  $3.528$ &  $5.084$ &  $3.719$ &  $4.595$ &  $\boldsymbol{3.517}$  \\
 noi-0-7-tel-tel       & $18.953$ &  $\boldsymbol{10.681}$ &  $15.518$ &  $11.276$ &  $15.868$ &  $12.280$ &  $\boldsymbol{8.831}$ &  $18.782$ &  $9.547$ &  $18.116$ &  $9.576$ &  $\boldsymbol{6.080}$ &  $11.402$ &  $6.252$ &  $10.014$ &  $6.382$  \\
 rev-noi-14-21-tel-tel & $16.517$ &  $15.099$ &  $11.044$ &  $11.356$ &  $9.398$ &  $\boldsymbol{7.631}$ &  $16.128$ &  $11.079$ &  $8.344$ &  $8.942$ &  $\boldsymbol{6.387}$ &  $6.379$ &  $6.097$ &  $4.798$ &  $4.948$ &  $\boldsymbol{4.143}$  \\
 rev-noi-7-14-tel-tel  & $19.543$ &  $19.899$ &  $13.985$ &  $15.615$ &  $12.352$ &  $\boldsymbol{9.610}$ &  $17.174$ &  $16.516$ &  $10.246$ &  $14.197$ &  $\boldsymbol{8.314}$ &  $7.169$ &  $8.184$ &  $5.626$ &  $7.033$ &  $\boldsymbol{5.126}$  \\
 rev-noi-0-7-tel-tel   & $27.834$ &  $28.154$ &  $22.193$ &  $24.523$ &  $21.442$ &  $\boldsymbol{16.841}$ &  $21.558$ &  $26.679$ &  $15.629$ &  $25.530$ &  $\boldsymbol{14.660}$ &  $10.533$ &  $15.609$ &  $8.770$ &  $14.369$ &  $\boldsymbol{8.149}$  \\
\bottomrule
\end{tabular}}
\end{table*}

\section{Experiments and discussion}
\label{exp}
We provide a set of results for answering two questions: (i) How does the speaker recognition performance depend on the type of the enhancement (denoising, dereverberation, both) and amount or type (real, artificial) of the autoencoder training data? (ii) How does using the autoencoder compare to using the multi-condition data for SRE system training? In the end we also combine the autoencoder with the multi-condition training and find the best performing combination.

We trained five different autoencoders for signal enhancement. Two autoencoders were trained only for dereverberation. The first was trained with artificially generated reverberation and the second used real reverberation. The third autoencoder was trained only for denoising. The last two autoencoders were trained simultaneously for denoising and dereverberation. Again, one of them used artificially generated RIRs and the second one used the real ones. 

Similarly, we created five different multi-condition training sets for PLDA. The approach is the same as in the autoencoder training. We used exactly the same noises and reverberation for segment corruption as in autoencoder training, allowing us to compare the performance when using the autoencoder or multi-condition training. 

Our results are listed in table \ref{tab:textindep}. 
Results are separated into two main blocks: PLDA trained on the clean data and PLDA trained on the multi-condition data. Each block is additionally separated to highlight whether the autoencoder enhancement is used or not. 

In the first block, the baseline corresponds to the system where the PLDA was trained only on the clean data without any enhancement. The next five columns represent results when using different autoencoders: N - autoencoder trained only on the noised data, AR - autoencoder trained on the data corrupted with artificial generated RIRs, RR-  autoencoder trained on the data corrupted with the real RIRs. N$+$(A/R)R - autoencoder simultaneously trained on the data with both types of distortion (noise and reverberation).

In the second block, we list the results for multi-condition training. We trained five different PLDAs, every time using a different mix of corrupted data added to the training list. PLDA or autoencoder on its own cannot fully profit from the added corrupted data. Autoencoder is able to partially remove the noise and reverberation from the data, while PLDA can learn the effect these data have for within- and across- speaker variability. Combining both techniques naturally brings the most improvement as we can see from the last block in table \ref{tab:textindep}. In these experiments, we were again modifying the data for the multi-condition PLDA training, but all of this data was previously processed by a single autoencoder. We decided to use the autoencoder simultaneously trained on the noisy and reverberated data (using real RIRs). This autoencoder was chosen based on its good and consistent performance in various conditions and we believe that it could represent an universal preprocessing step as there is only a negligible drop in performance when using it on clean data (see for example the performance on tel-tel condition of baseline system versus the N+RR column in the first block in table \ref{tab:textindep}).

Now, let us focus on comparing the baseline system and the system with enhanced data (PLDA is trained only on clean and enhanced data). In these experiments, we study which autoencoder training dataset is the best for given condition. 
If we look at these results globally, we can see that for most of the reverberation conditions (prism,rev, int-int, int-mic and rev-tel-tel, with exception of prism,chn), the autoencoder trained on the real reverberation provides the best results. Similar situation occurs for noisy conditions (prism,noi, noise-$\ast$-tel-tel) and noisy end reverberated conditions (rev-noise-$\ast$-tel-tel).  These results confirm our intuition, that it is best to use the autoencoder trained on the matching distortion to remove its effect from the data. We can also observe that to remove the reverberation, it is best to train on data reverberated by real RIRs instead of those artificially generated. This holds even for the condition containing only artificial reverberation (prism,rev). In general, when looking at the first block in table \ref{tab:textindep}, all of the autoencoders trained using reverberation with real RIRs (columns RR, N$+$RR) are better than those trained using artificial RIRs  (AR, N$+$AR).  We can also see, that the difference in performance between the RR-autoencoder and the N$+$RR autoencoder is rather small more in favor of the latter, both in reverberation and noisy conditions. This indicates that using the N$+$RR autoencoder is a good universal choice and justifies its selection for the experiments when combining the audio enhancing with multi-condition training.

When focusing on the multi-condition training (first part of the second block in table \ref{tab:textindep}) and taking the global view, we can observe similar trends as in the pure enhancement task. If we want to remove some type of distortion, it is best to add the matching distortion type into the PLDA training. If we look more closely, we can see the difference in reverberation conditions based on the PRISM set, where (as opposed to the enhancement) the multi-condition system using artificially generated RIRs have better results. This can indicate that it is easy for the PLDA to capture the channel variability caused by reverberating with the artificial RIRs which results in better performance in this matched-condition scenario. This hypothesis is further strengthened when comparing the AR with RR on rev-tel-tel condition when training on the matched-condition RR data almost halves the error rate.

If we analyze the difference in  performance between the pure signal enhancement and the multi-condition training, we see that the multi-condition training has slightly better results, especially in the hardest conditions rev-$\ast$-tel-tel. In the clean tel-tel condition, we can see that using autoencoder harms the performance less than multi-condition training. Additionally in some PRISM-based conditions (prism,rev, int-int, prism,chn), the autoencoder is also better than multi-condition training. 

Finally, we look at the combination of both techniques (the very last block in table \ref{tab:textindep}). Here, we are still having the same training lists for multi-condition PLDA training, but additionally, all data are enhanced by autoencoder trained on noised and reverberated data with real RIRs. We can see that in most conditions, we improve results with the pure multi-condition training. We suffer a significant degradation in clean tel-tel condition with respect to baseline for N+AR and N+RR training, but especially in the case of the latter, this degradation is compensated by excellent performance in other conditions, especially the most difficult rev-noise-$\ast$-tel-tel where we gain more than  $70\ \%$ relative improvement over the baseline. 

The combination of both techniques can also eliminate the big difference between artificially generated reverberation and real reverberation as can be seen by comparing results of  N$+$AG and N$+$RR systems. As we already saw for pure multi-condition training, the best results are again achieved by using the matched distortion for PLDA training, but the difference between the best possible results and multi-condition training with N$+$RR autoencoder are small. This justifies our recommendation to use the combination of multi-condition training with N$+$RR data that were preprocessed by the N$+$RR autoencoder as a universal and robust system, especially when expecting reverberated and/or noisy test data.

\section{Conclusion}
In this paper, we analyzed several aspects of DNN-autoencoder enhancement for designing robust speaker recognition systems. We studied the influence of different training sets on autoencoder performance in speaker recognition and we concluded that in our case the use of smaller amount of quality real RIRs provided better results than using much larger amount of artificial RIRs.

We also directly compared the PLDA multi-condition training with audio enhancing. Our results suggest that introducing the corrupted data on the i-vector level int the PLDA training provides slightly better results for noisy and reverberated condition, but at the same time causing more harm on clean data compared to the autoencoder.

Finally, we conclude that the combination of both techniques can significantly improve system performance compared to the baseline and even to systems using only one of the two techniques. We obtained more than $70\ \%$ relative improve with respect to baseline and approximately $40\ \%$ relative improvement with respect to multi-condition PLDA training. Based on our results and in the light of very good performance of MFCC-based systems in the NIST SRE 2016, we can say that autoencoders are a viable option to consider when designing a system that is robust against various levels of reverberation and noise.


\bibliographystyle{IEEEtran}

\bibliography{main}

\end{document}